\documentclass[eqsecnum,showpacs,aps,epsf,twocolumn,nofootinbib]{revtex4}
\usepackage{latexsym}
\usepackage{amssymb}
\usepackage{amsmath}

\newcommand{\bea}{\begin{eqnarray}}
\newcommand{\ena}{\end{eqnarray}}

\newcommand{\nn}{\nonumber\\}

\begin{document}

\title{The Kerr black hole and rotating black string by intersecting M-branes}

\author{Makoto Tanabe\footnote{e-mail address: tanabe@gravity.phys.waseda.ac.jp}}
\affiliation{Department of Physics Waseda University,
Ohkubo 3-4-1, Shinjuku-ku Tokyo 169-8555, Japan
}

\date{\today}

\begin{abstract}
We construct the non-BPS black brane representation from the Kerr metric using the U-duality and symmetries in string or M-theory. We give the general rule to get the brane configuration and we apply it especially for Kerr metric. We find the three charged solution in M-theory and after the compactification it becomes rotating black string solution in five dimension. We also find the four charged solution containing the pp-wave in M-theory, and we can find charged dilaton rotating black hole solution in four dimension after the torus compactification. This solution has the other representation in string theory, which is easy to apply the AdS/CFT correspondence. 
\end{abstract}

\maketitle

\section{Introduction}
In recent years, many varieties of five-dimensional vacuum solutions have been founded \cite{BR,2BR,ST,CBR,BS,BC,BdiRing} using the rod structure formalism \cite{rod,SolGen}. It is impossible to get the charged black hole solutions using the rod structure formalism \cite{BoostBrane,BoundMBrane}, thus we use the other method to get the charged solutions using the U-duality \cite{Tdual} in the super string theory. This method makes the non-BPS solutions of the string theory \cite{Udual}, and especially this method succeed to find the non-BPS black ring solution from the one rotating vacuum black ring solution \cite{NONBPSBR}. We must add the additional dipole charge for the ordinary solution to get the seed of another rotating axis. Thus we give the general formalism to get the charged solutions using the duality in string or M-theory without adding any dipole charge in ordinary seed metric.

Non BPS black ring solutions \cite{NONBPSBR} using the same method we use below, but these solutions related to intersecting $M2\perp M2\perp M2$-brane solution in M-theory, or intersecting $F1\perp D2\perp D2$-brane solution in type IIA string theory with compactification on one-dimensional torus of M-Theory. These configuration in string theory is impossible to analysis for AdS/CFT correspondence \cite{AdS/CFT}, because we must choose typical coordinates for the compact space for near horizon limit, which is the kk-wave direction, but this solution has not any typical direction for compact spaces. 

Thus we introduce the $D1\perp D5$-brane solutions from general axisymmetric five-dimensional vacuum solution with three Killing vector, which contain all solutions we already know from the rod structure. We show the $NS5 \perp D4\perp D4$-brane solutions from general axisymmetric four-dimensional vacuum solution with two Killing vector, e.g., Kerr metric. We also get the general rotating black string solutions, which is rotating four-dimensional black hole with extra one dimension, by the $M2\perp M2\perp M2$-brane solutions from the Kerr metric.

\section{Charging up via U-duality}

We introduce the method for charging up from the vacuum solution, which we use the boost, T-duality and S-duality \cite{Tdual}. The vacuum solution satisfy the vacuum Einstein equation $G_{\mu\nu}=0$, and if we add the extra flat dimension, this still satisfies the same equation as $G_{MN}=0$, where 
$x^M$ is the higher dimensional coordinate. Therefore the Higher dimensional solutions satisfy the Einstein equation, and obviously there are no Maxwell field or gauge field. We consider the ten-dimensional supergravity below, because this theory has the T-duality and S-duality \cite{Udual}. 

Under the boost for these solutions, it is still the solution because the all super gravity theory has the Lorentz symmetry, then the solutions change to wave solution from the vacuum solution. We denote this boost process as $B_{\alpha}(x^\mu)$, which means that boosted for the $x^\mu$ direction with the boost parameter $\alpha$. 

T-duality is the symmetry between compact radius $R \leftrightarrow 1/R$,  and it interchange to the type IIA and IIB theory. We use the notation for the T-duality along $x^\mu$ direction as $T(x^\mu)$. Under the T-duality for the direction of the wave, the solution changes to a fundamental string $F1$-brane, and $Dp$-brane change to $D(p-1)$-brane when the T-duality direction is parallel to the $Dp$-brane direction, and otherwise becomes $D(p+1)$-brane. 

S-duality is the symmetry between the strong/weak coupling, and type IIB theory is self-dual under the S-duality, and $F1$-brane interchanges to the $D1$-brane and $NS5$-brane interchange to the $D5$-brane. 

We must also consider the Hodge dual for the antisymmetric gauge field, and specially D3-brane are self-dual field, thus the field strength of $D3$-brane and its dual field strength are the same. 

When we lift up to the eleven dimension, we find the M-theory which is strong coupling limit of type IIA supergravity, and M-theory contains only $M2$ and $M5$-brane. $M2$-brane comes from the $F1$ and $D2$-brane, and $M5$-brane comes from the $D4$ and $NS5$-brane. In eleven dimension there is no symmetry under the T-duality and S-duality, but still the Lorentz symmetry exists, thus we can only apply the boost in M-theory. 

\section{From Kerr metric to Intersecting D-branes}

In four dimension we have a unique rotating vacuum solution as the Kerr metric, which is given by 
\begin{eqnarray*}
ds^2=-f\left(dt+\Omega \right)^2 
+\Sigma^2 \left({dr^2\over \Delta}+d\theta^2\right)
+{\Delta\over f}\sin^2\theta d\phi^2 
\end{eqnarray*}
where the metric functions are defined by 
\begin{eqnarray*}
&&\Sigma^2=r^2+a^2\cos^2\theta
\,,~ \Delta= r^2-2mr+a^2
\nn
&&\Omega={2mr \over \Sigma^2 -2mr}a\sin^2\theta d\phi
\,,~
f=1-{2mr \over \Sigma^2}
\,.
\end{eqnarray*}
The mass of the black hole is $m$ and $a$ is the specific angular momentum witch bounded for $m\geq a$. 

Now we generalize the metric as $ds^2=-f (dt+\Omega)^2+ds_{\rm base}^2$, where $\Omega=\Omega_\phi(r,\theta) d\phi$ are the arbitrary function and $ds_{\rm base}^2=\gamma_{ij}dx^idx^j$ are the orthogonal three-dimensional metric written by the variables $r,\theta, \phi$. The metric has the two Killing vector $\xi_t$ and $\xi_\phi$. We add the other six flat dimensions and we apply the sequence; 
\begin{eqnarray*}
B_{\alpha_2}(z_1)\rightarrow 
T(z_1)\rightarrow S \rightarrow 
T(z_2)\rightarrow T(z_3) 
\,,
\end{eqnarray*}
then we find the $D3$-brane solutions in type IIB theory. The metric of $D3$-brane is given by
\begin{eqnarray}
&&ds^2=-h_2^{-1/2}f (dt+c_2\Omega)^2
+h_2^{1/2}ds_{\rm base}^2
\nn
&&~~~~~~~~~~~
+h_2^{1/2}\left[h_2^{-1}\sum_{i=1}^3dz_i^2
+\sum_{i=4}^6 dz_i^2\right]
\,,
\end{eqnarray}
where the metric function are given by $h_2=-s_2^2 f+c_2^2$ and the dilaton field is given by $e^{-2\varphi}=1$. The gauge fields related to $D3$-brane are given by the four-form fields as 
$4{\tilde D}_{z_1z_2z_3 t}= h_2^{-1}(f-1)s_2c_2$ and $4{\tilde D}_{z_1z_2z_3 \phi}=h_2^{-1}fs_2 \Omega_\phi$, where we use the notation as $\cosh \alpha_i =c_i$ and $\sinh \alpha_i=s_i$ for the simplicity. To continue the sequence we must calculate the Hodge dual as 
\begin{eqnarray}
&&\partial_r {\tilde D}_{z_4 z_5 z_6 t}=h_2  f
{\gamma_{rr}\over \sqrt{-\gamma}}
\left[c_2\Omega_\phi \partial_\theta D_{z_1z_2z_3 t}
-\partial_\theta D_{z_1z_2z_3 \phi}\right]
\nn
&&\partial_r {\tilde D}_{ z_4 z_5 z_6 \phi}=h_2 f
{\gamma_{rr}\over \sqrt{-\gamma}}
\left[c_2\Omega_\phi \partial_\theta D_{z_1z_2z_3 \phi}
\right.
\nn
&&\left. ~~~~~~~~~~~~~~
+(-c_2^2\Omega_\phi^2+h_2f^{-1}\gamma_{\phi\phi})
\partial_\theta D_{z_1z_2z_3 t}\right]
\label{eqn:Hodge1}
\,,
\end{eqnarray}
and  $\theta$ derivative components are the same as this. We denote ${\tilde D}_{z_4z_5z_6 t}=s_2 {\tilde D}_t$ and ${\tilde D}_{z_4z_5z_6 \phi}=s_2c_2 {\tilde D}_\phi$ in below.

In Kerr metric case, the Hodge dual can be written in  
\begin{eqnarray}
&&\partial_r  4{\tilde D}_\phi =- f^2 
{\gamma_{rr}\over \sqrt{-\gamma}}
\left( \Omega_\phi \partial_\theta \Omega_\phi 
+\gamma_{\phi\phi}f^{-2}\partial_\theta f\right)
\nn
&&\partial_r  4{\tilde D}_t = f^2 
{\gamma_{rr}\over \sqrt{-\gamma}}\partial_\theta \Omega_\phi 
\,,
\end{eqnarray} 
thus we can calculate the dual fields as 
\begin{eqnarray}
4{\tilde D}_t= {ma \cos\theta \over \Sigma^2}
\,,~
4{\tilde D}_\phi= {ma^2 \cos\theta \sin^2\theta \over \Sigma^2}
\,.
\end{eqnarray}

This sequence continues as follows; 
\begin{eqnarray*}
&&B_{\alpha_1}(z_1)\rightarrow T(z_1) \rightarrow S
\rightarrow T(z_5)\rightarrow T(z_6)\rightarrow T(z_2)
\,,
\end{eqnarray*}
and we get the $D4\perp D4$-brane solutions as 
\begin{eqnarray}
&&ds^2=-(h_1h_2)^{-1/2}f(dt+c_1c_2\Omega)^2
+(h_1h_2)^{1/2}ds_{\rm base}^2
\nn 
&&~~~
+(h_1h_2)^{1/2}\left[
h_1^{-1}(dz_1^2+dz_2^2)+ h_2^{-1}( dz_3^2 +dz_4^2) \right]
\nn
&&~~~
+(h_1h_2)^{-1/2}g_3^{-1}(dz_5^2+dz_6^2)
\,,
\end{eqnarray}
where $h_1=-s_1^2f+c_1^2$ and $g_3=1-(h_1h_2)^{-1}(4s_1 s_2 {\tilde D}_t)^2$. The dilaton field becomes $e^{-2\varphi}=(h_1h_2)^{1/2}g_3$ and the gauge fields change as below: 
\begin{eqnarray}
&&B_{z_5z_6}=(h_1h_2g_3)^{-1}4s_1 s_2 {\tilde D}_t 
\nn
&&{\tilde C}_{z_1z_2 t}={8\over 3}h_1^{-1}c_1 s_2{\tilde D}_t \,,
\nn
&&{\tilde C}_{z_1z_2 \phi}={8\over 3}s_2c_2
 \left[ {\tilde D}_\phi-s_1^2 h_1^{-1}  {\tilde D}_t f  \Omega_\phi\right] \,,
\nn
&&{\tilde C}_{z_3z_4 t}={8\over 3}h_2^{-1}s_1 c_2{\tilde D}_t \,,
\nn
&&{\tilde C}_{z_3z_4 \phi}={8\over 3}s_1c_1
 \left[ {\tilde D}_\phi-s_2^2 h_2^{-1}  {\tilde D}_t f  \Omega_\phi\right] 
\,.
\end{eqnarray}

For simplify we change the notation as
\begin{eqnarray*}
&&{\bar D}^{(1)}_t= h_1^{-1}c_1 s_2 {\tilde D}_t 
\,, ~ 
{\bar D}^{(1)}_\phi=s_2c_2\left( {\tilde D}_\phi
-s_1^2 h_1^{-1}  {\tilde D}_t f  \Omega_\phi\right)
\nn
&&{\bar D}^{(2)}_t= h_2^{-1}s_1 c_2 {\tilde D}_t 
\,, ~ 
{\bar D}^{(2)}_\phi=s_1c_1 \left( {\tilde D}_\phi
-s_2^2 h_2^{-1}  {\tilde D}_t f  \Omega_\phi\right)
\,.
\end{eqnarray*}

We can also calculate for the Kerr metric, then we find 
\begin{eqnarray}
&&{\bar D}^{(1)}_t=h_1^{-1}c_1 s_2{\tilde D}_t 
\,,~
{\bar D}^{(1)}_\phi =h_1^{-1}s_2c_2{\tilde D}_\phi  
\nn
&&{\bar D}^{(2)}_t=h_2^{-1}s_1 c_2{\tilde D}_t 
\,,~
{\bar D}^{(2)}_\phi =h_2^{-1}s_1c_1{\tilde D}_\phi  
\,.
\end{eqnarray} 

We apply the next sequence to add another charge as 
\begin{eqnarray*}
B(z_6)\rightarrow T(z_6) \rightarrow S \rightarrow T(z_5)
\,,
\end{eqnarray*} 
we find $D2\perp D2\perp D2$ brane solution as 
\begin{eqnarray*}
&&ds^2=-\xi^{-1/2}f(dt+c_1c_2c_3\Omega)^2 
+\xi^{1/2}ds_{\rm base}^2
\nn 
&&~~~
+\xi^{1/2}\left[h_1^{-1}\sum_{i=1}^2dz_i^2
+ h_2^{-1}\sum_{i=3}^4 dz_i^2 
+ h_3^{-1}\sum _{i=5}^6 dz_i^2\right] 
\label{eqn:d2d2d2}
\,,
\end{eqnarray*}
with the new function $h_3=-s_3^2 f+c_3^2$ and $\xi=h_1h_2{\hat h_3}g_3=h_1h_2h_3 -\beta_t^2$ where ${\hat h}_3=-fs_3^2+g_3^{-1}c_3^2$ and $\beta_t=4 s_1s_2 s_3 {\tilde D}_t$. The dilaton field is $e^{-2\varphi}= \xi^{-3/2}h_1h_2h_3$ and the gauge field is  
\begin{eqnarray*}
&&B_{z_iz_j}=h_\alpha^{-1}{c_\alpha \over s_\alpha} \beta_t 
\,, ~ 
{\tilde A}_t=-\xi^{-1}\beta_t f
\,,~
{\tilde A}_\phi =-\xi^{-1}\beta_t f \omega
\nn
&&{\tilde C}_{z_iz_j t}={2\over 3}h_\alpha^{-1}s_\alpha c_\alpha  (f-1)
\,, ~ 
{\tilde C}_{z_i z_j\phi}
={2\over 3} h_\alpha^{-1}fc_\alpha^{-1}s_\alpha\omega_\phi 
\,.
\end{eqnarray*}
where the pair of indices $(i,j)=(1,2),(3,4),(5,6)$ are corresponding to $\alpha=1,2,3$. 
For the first example of a black hole solution, we continue to apply the U-duality for the charging up the four-dimensional Kerr solutions. Next we try to apply another example related to the rotating black string, which can be described by the Kerr metric with another one extra dimension, and these solutions must possess the Gregory-Laflamme instability \cite{GLinst}.

\subsection{Kerr solutions in String/M-theory}

For the adding pp-wave in eleven dimension, we must apply the sequence as 
\begin{eqnarray*}
T(z_1)\rightarrow T(z_3) \rightarrow T(z_5) 
\rightarrow T(z_2) \rightarrow T(z_4) \rightarrow T(z_6)
\,,
\end{eqnarray*}
then we find the $D4\perp D4 \perp D4$-brane solutions as  
\begin{eqnarray*}
&&ds^2=\xi^{1/2}\left[{\bar h}_1^{-1}\sum_{i=1}^2dz_i^2
+ {\bar h}_2^{-1}\sum_{i=3}^4dz_i^2
+ {\bar h}_3^{-1}\sum_{i=5}^6dz_i^2
\right] 
\nn 
&&~~~~
-\xi^{-1/2}f(dt+c_2c_1c_3\Omega)^2 
+\xi^{1/2}ds_{\rm base}^2
\,,
\end{eqnarray*}
where ${\bar h}_i=\xi {\hat h}_i^{-1}$ with ${\hat h}_i =-s_i^2 f +c_i^2 g_i^{-1}$. The function $g_i$ is determined by $g_i= 1+(h_1h_2h_3)^{-1}h_i^{-1}s_2^{-2}\beta_t^2$. The dilaton field is $e^{-2\varphi}=\xi^{-3/2}{\bar h}_1 {\bar h}_2 {\bar h}_3$ and the gauge fields are given by ${\tilde A}_a={\hat A}_a$ and  
\begin{eqnarray*}
&&B_{z_iz_j}={\bar h}_\alpha^{-1}{c_\alpha \over s_\alpha} \beta_t 
\,,~ 
{\tilde C}_{z_iz_ja}
={8\over 3} {\hat D}^{(\alpha)}_a
+{\bar h}_\alpha^{-1}s_\alpha^{-1}c_\alpha \beta_t {\hat A}_a
\,,
\end{eqnarray*}
where we use the same notation for $(i,j)$ and the Hodge dual fields ${\hat D}^{(\alpha)}_a$ can be calculate as 
\begin{eqnarray*}
&& {\hat D}_t^{(1)}
=-\xi^{-1}h_2 s_1c_2c_3 {\tilde D}_t 
\nn 
&& {\hat D}_\phi^{(1)}
=s_2c_2 \left[ {\tilde D}_\phi 
+(1-\xi^{-1}h_2 c_1^2 c_3^2) {\tilde D}_t \Omega_\phi  \right]
\nn 
&& {\hat D}_t^{(2)}
=-\xi^{-1}h_1 c_1s_2c_3 {\tilde D}_t 
\nn 
&& {\hat D}_\phi^{(2)}
=s_1c_1 \left[ {\tilde D}_\phi 
+(1-\xi^{-1}h_1 c_2^2 c_3^2 ){\tilde D}_t \Omega_\phi \right]
\nn 
&& {\hat D}_t^{(3)}
=-\xi^{-1}h_3 c_1c_2s_3 {\tilde D}_t 
\nn 
&& {\hat D}_\phi^{(3)}
=s_3c_3 \left[ {\tilde D}_\phi 
+(1-\xi^{-1}h_3 c_1^2 c_2^2) {\tilde D}_t \Omega_\phi  \right]
\,,
\end{eqnarray*}
The function ${\hat A}_a$ is also the Hodge dual fields as   
\begin{eqnarray}
&&\partial_r {\hat A}_\phi 
= {\gamma_{rr}\over \sqrt{-\gamma}}\beta_t 
\left[f^2 \omega_\phi \partial_\theta \omega_\phi
+\xi^2 \beta_t^{-1}\gamma_{\phi\phi}\partial_\theta (\xi^{-1}\beta_t f)\right]
\nn
&&\partial_r {\hat A}_t
= {\gamma_{rr}\over \sqrt{-\gamma}}
\beta_t f^2 \partial_\theta \omega_\phi
\,,
\label{eqn:dual3}
\end{eqnarray}
and of cause we can calculate the $\theta$ derivative in the same way. In Kerr metric we find that ${\hat A}_a$ can be written by 
\begin{eqnarray*}
&&{\hat A}_t= s_1c_1s_2c_2 s_3c_3 {ma^2 \cos^2\theta \over \Sigma^4}
\nn
&&{\hat A}_\phi =-s_1s_2s_3 \left (c_1^2c_2^2c_3^2 
+{2r \Sigma^2 h_1h_2h_3 \over r^2-a^2\cos^2\theta }\right)
{ma^3 \sin^2\theta \cos^2\theta \over \Sigma^4}
\,.
\end{eqnarray*}

Now we lift up the $z^7$ direction then we find the $M5\perp M5\perp M5$-brane with pp-wave solutions in M-theory; 
\begin{eqnarray}
&&ds^2=\Xi^{1/3}
\left[{\bar h}_1^{-1}\sum_{i=1}^2dz_i^2
+ {\bar h}_2^{-1}\sum_{i=3}^4dz_i^2
+ {\bar h}_3^{-1}\sum_{i=5}^6dz_i^2
\right] 
\nn 
&&~~~~
+\Xi^{1/3}\left[ 
-\xi^{-1}f(dt+\omega d\phi)^2 
+ds_{\rm base}^2\right]
\nn 
&&~~~~
+\Xi^{-2/3}\xi\left(dz_7+{\hat A}_tdt+{\hat A}_\phi d\phi\right)^2 
\,,
\end{eqnarray}
where the conformal factor can be written by $\Xi= {\bar h}_1{\bar h}_2{\bar h}_3$. The three-form fields related to $M5$-brane are given by 
\begin{eqnarray*}
&&{\hat C}_7^{(\alpha)}\equiv 
C_{z_iz_jz_7}={2\over 3}{\bar h}_\alpha^{-1}s_\alpha^{-1}c_\alpha \beta_t 
\nn
&&{\hat C}_a^{(\alpha )}\equiv 
{\tilde C}_{z_iz_ja}
={8\over 3}{\hat D}^{(\alpha)}_a
+{\bar h}_\alpha^{-1}s_\alpha^{-1}c_\alpha \beta_t {\hat A}_a
\,,
\end{eqnarray*}
We note that Kerr metric with flat dimensions has no Chern-Simons term, but there are exist in eleven dimension after the charging up sequence, e.g., ${\hat C}^{(1)}_7 \wedge d{\hat C}^{(2)}_t \wedge {\hat C}^{(3)}_\phi $ or like that combinations. The Chern-Simons terms gives the non-trivial effect of the topology for BPS black ring solutions \cite{SUSYBR}, and this effect change the Laplace equation for the harmonic function $h_i$ to the Poisson equation for the non-harmonic function ${\bar h}_i$.

Finally we apply the boost for the $z_7$ direction, and compactify the $z_6$ direction to apply the AdS/CFT correspondence, we find $NS5\perp D4\perp D4$-brane solutions with kk-wave, and the metric can be written by 
\begin{eqnarray}
&&ds^2=( {\bar h}_1 {\bar h}_2)^{1/2}
\left[{\bar h}_1^{-1}\sum_{i=1}^2dz_i^2
+ {\bar h}_2^{-1}\sum_{i=3}^4dz_i^2
+ {\bar h}_3^{-1}dz_5^2
\right] 
\nn 
&&~~~~
+( {\bar h}_1 {\bar h}_2)^{1/2}\left[ -\xi^{-1}fdu^2 +\Xi^{-1}\xi dv^2 
+ds_{\rm base}^2\right]
\,,
\end{eqnarray}
where $du$ and $dv$ are determined by 
\begin{eqnarray*}
&&du=c_4dt+c_2c_1c_3\Omega+s_4 dz^7
\nn
&&dv=(s_4+c_4{\hat A}_t)dt 
+{\hat A}_\phi d\phi+(c_4+s_4{\hat A}_t)dz^7
\,,
\end{eqnarray*}
The dilaton field becomes $e^{-2\varphi}=\xi^{-1/2}{\bar h}_3^{3/2}$, and the gauge fields are given by and the three-form fields are given by 
\begin{eqnarray*}
&&{\tilde C}_{z_iz_j t}=c_4{\hat C}^{(\alpha )}_t +s_4{\hat C}^{(\alpha)}_7 
\,,~
B_{z_5t }=-{3\over 2}\left(c_4{\hat C}^{(3)}_t +s_4{\hat C}^{(3)}_7 \right)
\nn
&&{\tilde C}_{z_iz_jz_7}=c_4{\hat C}^{(\alpha)}_7 +s_4{\hat C}^{(\alpha)}_t 
\,,~
B_{z_5z_7}=-{3\over 2}\left(c_4{\hat C}^{(3)}_7 +s_4{\hat C}^{(3)}_t\right)  
\nn
&&{\tilde C}_{z_iz_j\phi}={\hat C}^{(\alpha)}_\phi 
\,,~
B_{z_5z_6\phi}={3\over 2}{\hat C}^{(3)}_\phi 
\,,
\end{eqnarray*}
where the indices $(i,j)$ are only run in $(1,2)$ and $(3,4)$. 

After the compactify the other six dimension on torus, we find the four dimensional charged solution as 
\begin{eqnarray}
&&ds^2=-\Upsilon f(dt+{\bar \omega} d\phi )^2 
+\Upsilon^{-1} ds_{\rm base}^2
\label{eqn:ChargedKerr}
\,,
\end{eqnarray}
where ${\bar \omega}$ and $\Upsilon$ are determined by 
\begin{eqnarray*}
&&{\bar \omega}=c_4 \omega +s_4 (\omega {\hat A}_t-{\hat A}_\phi )
\nn
&& \Upsilon^{-2}=\Xi \left(
 -\xi^{-1}fs_4^2+\Xi^{-1}\xi (c_4+s_4 {\hat A}_t)^2\right)
\,,
\end{eqnarray*}
This metric must be satisfied  the equation of motion which is given by the effective sigma model action in four dimension as 
\begin{eqnarray*}
&&S_4=\int d^4 x {\sqrt{-g}\over 16\pi G_4}
\left[ {\cal R}_4 + \sum_{\alpha=1}^3 \left|\nabla \phi_\alpha\right|^2 
+{1\over 2} \left|\nabla \phi_4\right|^2 
\right.\nn 
&&~~~~~~~~~~~~~\left. 
+\sum_{\alpha=1}^3 e^{-2\phi_\alpha}
 \left|\nabla \rho_\alpha\right|^2 
+\sum_{\alpha=1}^4 e^{2(\varphi -\phi_\alpha)}\left |d{\cal A}_a^{(\alpha)}\right|^2 \right] 
\,,
\end{eqnarray*}
 where ${\cal R}_4$ is the four dimensional Ricci scalar  determined by the metric \eqref{eqn:ChargedKerr} and   $\varphi= 2\phi_1+2\phi_2 +2\phi_3+\phi_4$. The scalar field and the gauge fields are given  by 
 \begin{eqnarray}
&&e^{2\phi_\alpha}=\Xi^{1/3}{\bar h}_{i}^{-1}
 \,, ~
 e^{2\rho_\alpha}\sim  c_4 {\hat C}^{(\alpha)}_7+s_4 {\hat C}^{(\alpha)}_t
 \nn
 &&{\cal A}^{(\alpha )}_t\sim c_4 {\hat C}^{(\alpha)}_t+s_4 {\hat C}^{(\alpha)}_7
 \,, ~ 
 {\cal A}^{(\alpha)}_\phi \sim {\hat C}^{(\alpha)}_\phi 
 \nn 
 &&e^{2\phi_4}=\Xi^{-2/3}\xi 
 \,,~
 {\cal A}^{(4)}_a={\hat A}_a
\,.
\label{eqn:Kerrgauge}
 \end{eqnarray}
In the Kerr metric case, the regularity condition for the rotating axis are the same as before, thus the metric has no conical singularity at the ordinary event horizon $r_+=m+\sqrt{m^2-a^2}$.

In the asymptotic region $(r\rightarrow \infty)$ the metric becomes flat and the ADM mass $M=m(1+\sum_{i=1}^4s_i^2/2)$, the conserved charge $Q=\sqrt{2}=\sqrt{2}m\sum_{i=1}^4 s_ic_i/2$ and the angular momentum $J=c_1c_2c_3c_4 a$ are given in the asymptotic metric form. The surface gravity change as below 
\begin{eqnarray}
\kappa={1 \over \beta_{t+}^2 (c_4+s_4 {\hat A}_{t+})} 
{r_+^2-a^2 \over 4mr_+^2}
\,,
\end{eqnarray}
where $\beta_{t+}$ and ${\hat A}_{t+}$ are defined by the substitution for $r=r_+$. The surface area of the outer event horizon 
\begin{eqnarray*}
&&{\cal A}=\int d\theta d\phi \left.
\sqrt{\Sigma^2 \left(\Upsilon^{-2}\gamma_{\phi\phi}
-f{\bar \omega}^2\right)}
 \right|_{r=r_+}
\nn
&&~~
=8\pi mr_+ c_1c_2c_3\left[ c_4
-{1\over 2} a^3 s_4 \left(1-{a\over r_+}{\rm arctan}~ {r_+\over a} \right)\right]
\,,
\end{eqnarray*}
and we can show the thermodynamics with the physical parameter as the charge and angular momentum and the temperature, but the dilaton fields does not contribute.

 Sen gave the rotating charged black hole solution \cite{SenBH}, which metric in Einstein flame is given by 
 \begin{eqnarray}
&&ds^2=- h_\alpha^{-1} f\left( dt+c_\alpha^2 \Omega\right)^2 
+h_\alpha ds_{\rm base}^2
\,, 
\end{eqnarray}  
where $h_\alpha=-s_\alpha^2 f+c_\alpha^2$. Sen's solution is included in our solution with the parameter 
$c_1=c_2=c_\alpha$ and $c_3=c_4=1$, and this case the action is changing as $\phi_1=\phi_2$ and $\phi_3=\phi_4=0$ and $\rho_\alpha=0$ and only ${\cal A}^{(1)}_a={\cal A}^{(2)}_a$ are exist.

The static limit for $a=0$ we have ${\tilde D}_t=0$ and ${\hat A}_a=0$, thus the metric becomes four charged static black brane solutions as 
\begin{eqnarray*}
ds^2=-h^{-1/2}f dt^2 +h^{1/2}\left(f^{-1}dr^2 +r^2d\Omega^2\right)
\,,
\end{eqnarray*}
where $h=h_1h_2h_3h_4$. The harmonic function $h_i$ defined as $h_i=-s_i^2 f+c_i^2$ with $f=1-2m/r$, which is the metric function of ordinary Shwarzshild metric. This solutions are contain the Gibbons-Maeda's dilatonic black hole solutions \cite{GibbonsMaeda}. 

We can also take the extremal limit for the Kerr metric ($a=m$), this case the horizon are degenerate and it will contain the supersymmetric solution, because these metric configuration in M-theory are the same figure as we find before \cite{SSfromIM}.

We have coordinate singularity at $\xi=0$, witch is the same as ${\bar h}_i=0$, and at this case the Kretchmann invariance $R_{\mu\nu\rho\sigma}R^{\mu\nu\rho\sigma}$ is diverging. If we choose the specific value for the $\alpha_i$, $m$ and $a$, we can get the regular solitonic solution, and we will show in the next paper.

\subsection{Rotating Black String Solution}

Now we apply the other example about charging up method from four-dimensional vacuum solution, and it can be applied for black string in five dimension, which is the Kerr metric with the extra one dimension.  
We lift up the $z^7$ direction directory for \eqref{eqn:d2d2d2}, then we find $M2\perp M2\perp M2$-brane solution as 
\begin{eqnarray}
&&ds^2=\Theta^{1/3}
 \left[h_1^{-1}\sum_{i=1}^2dz_i^2
+ h_2^{-1}\sum_{i=3}^4dz_i^2
+ h_3^{-1}\sum_{i=5}^6dz_i^2
\right] 
\nn 
&&~~~~
+\Theta^{1/3}\left[ -\xi^{-1}f (dt+\omega d\phi)^2 
+ds_{\rm base}^2\right]
\nn
&&~~~~
+\Theta^{-2/3}\xi [-\xi^{-1}\beta_tf (dt+\omega d\phi)+dz_7]^2
\,,
\end{eqnarray}
where the conformal factor is $\Theta=h_1h_2h_3$, and the three-form fields related to $M2$-brane are given by 
\begin{eqnarray*}
&&C_{z_iz_jz_7}={2\over 3}h_\alpha^{-1}{c_\alpha \over s_\alpha } \beta_t 
\nn
&&C_{z_iz_j t}={2\over 3}h_\alpha^{-1}s_\alpha c_\alpha (f-1)
\nn
&&C_{z_i z_j\phi}
={2\over 3} h_\alpha^{-1}f{s_\alpha \over c_\alpha} \omega_\phi 
\,.
\end{eqnarray*}
All metric function and the gauge field are given by the harmonic function $h_i$ and the extra function $\beta_t$, which comes from the Hodge dual for the gauge field. The Chern-Simons terms are also exist in this case, however if we take that the extra function $\beta_t$ must be zero, then the Chern-Simons terms are vanishing and all metric function only depend on the harmonic function $h_i$. This gives the well known solutions given by \cite{SSfromIM}, but in this case one of the function $h_i$ must be zero, thus we only take the two charged solution with Kerr based metric. 

Compactify the $z_1$ to $z_6$ direction we find charged rotating black string solutions as 
\begin{eqnarray}
&&ds^2=\Theta^{1/3}\left[ -\xi^{-1}f (dt+\omega d\phi )^2 
+ds_{\rm base}^2 \right] 
\nn 
&&~~~~
+\Theta^{-2/3}\xi \left[-\xi^{-1}\beta_t f 
(dt+\omega d\phi)+dz_7\right]^2
\label{eqn:black string}
\,.
\end{eqnarray}

This solutions satisfy the effective sigma model action after the compactification on tori as 
\begin{eqnarray*}
&&S_5=\int d^5 x {\sqrt{-g}\over 16\pi G_5}
\left[ {\cal R}_5 + \sum_{\alpha=1}^3 \left|\nabla \phi_\alpha\right|^2 
+{1\over 2} \left|\nabla \varphi\right|^2 
\right.\nn 
&&~~~~~~~~~~~~~\left. 
+\sum_{\alpha=1}^3 e^{2(\varphi-\phi_\alpha)}\left |d{\cal A}_a^{(\alpha)}\right|^2 \right] 
\,,
\end{eqnarray*}
where ${\cal R}_5$ is the five dimensional Ricci scalar given by the metric \eqref{eqn:black string}. The scalar field $\phi_\alpha$ are the same as before 
\eqref{eqn:Kerrgauge}, and $\varphi=\sum_{\alpha=1}^3 \phi_\alpha/3$. 
The gauge field is given by 
${\cal A}^{(\alpha)}_t=c_4 {\hat C}^{(\alpha)}_t+s_4 {\hat C}^{(\alpha)}_7
\,,~
{\cal A}^{(\alpha)}_7=c_4 {\hat C}^{(\alpha)}_7+s_4{\hat C}^{(\alpha)}_t $
and ${\cal A}^{(\alpha)}_\phi ={\hat C}^{(\alpha)}_\phi$.

The horizon of the solutions are the same as the ordinary one, and this case regularity condition for the rotating axis $\phi$ are the same as before. This solution has the new singularity at $\xi=0$ or $\beta_t\rightarrow \infty$, but $\xi=0$ are satisfied both $h_i=0$ and $\beta_t=0$, which conditions give trivial solution, unless $f=c_i^2/s_i^2$. 

When we fix the extra coordinate $z^7=z^7_0$, then we find the metric becomes 
\begin{eqnarray}
ds^2=\Theta^{-2/3}f(dt+\omega d\phi)+\Theta^{1/3}ds_{\rm base}^2
\,,
\end{eqnarray}
and this metric form is the consistent as the solution from supersymmetric five dimensional solution. 
The area surface for a unit length of $z_7$ direction, we can calculate in generally, but it is very complicated figure. In this time we show the special case as $s_1=s_2=s_3\equiv s$ case, then we find the surface area as 
\begin{eqnarray}
&&{\cal A}=\int d\theta d\phi \sqrt{\Theta ^{-1/3}\Sigma^2 f \omega^2}
\nn 
&&~~
=4\pi (r_+^2+a^2) \left[1+\left({r_+^2 \over a^2}-\alpha^2\right)
{{\rm Arccot} ~\alpha \over \alpha}\right]
\,,
\end{eqnarray}
where $\alpha^2 a^2=r_+^2-2mr_+s^2$. In the static limit, the area surface is equal to one of the Schwarzshild black string. We can easy to check the non-charged solution is given by $s=0$, and  we find the area surface ${\cal A}=4\pi (r_+^2+a^2)$, which is consistent with the one of ordinary Kerr metric.

In the Kerr metric at the infinity the conformal factor becomes $\Theta \rightarrow 1$, thus the metric becomes Minkowski metric, which satisfy the asymptotically flat. The mass, angular momentum and the charge of the Maxwell fields in this metric change to $M=m(s_1^2+s_2^2+s_3^2)/3$, $J=Ma$ and $Q=m (s_1c_1+s_2c_2+s_3c_3)/3$. This solution is obviously different from the Kerr-Newmann metric, which is only for the changing the $\Delta=r^2-2mr+a^2+q^2$ and $f=1-(2mr-q^2)/\Sigma$. The action in four dimension is non-minimal coupling to the dilaton with the Maxwell field, thus we are not allow to take the limit just for vanishing the dilaton, only we can take the limit for no dilaton and no Maxwell fields, which is related to $s_i=0$. 

In the static limit $a=0$, the metric becomes 
\begin{eqnarray}
ds^2=-\Theta^{-1}fdt^2 +f^{-1} dr^2 + r^2 d\Omega^2 +\Theta dz_7^2
\,,
\end{eqnarray}
 where the metric function change as $f=1-2m/r$. The dilaton fields are depend only on the radial coordinate $r$, and the gauge fields becomes only electrical potential $A_t=\sum h_\alpha^{-1}s_\alpha c_\alpha (f-1)/3$, and we can easy to check the charge is $Q=m(s_1c_1+s_2c_2+s_3c_3)/3$, which means that charge is independent for the angular momentum. Both of the static and stationary case, the dilaton fields also do not contribute the thermodynamics of black string solution for the unit per length along the extra dimension. 
  
  This solution has the instability of the extra dimension along $z^7$ direction, named Gregory-Lafflame Instability. In the static case, we can calculate in the general way in \cite{GLreview}, but in the stationary case the perturbation of the metric function are impossible to separable for the variables. Thus we must concern about the  analyze about it.

\section{Concluding remark}
In this paper we have presented the charging up method from the four dimensional vacuum solution with rotating axis, Kerr metric. We use the U-duality and Lorentz symmetries in higher dimensional supergravity theory, i.e., ten-dimensional type IIA and IIB supergravity, which is the low energy limit of the type IIA and IIB super string theory, and eleven-dimensional supergravity, named M-theory. 

We give a specific example as charging up the Kerr solution in four dimension, which relates the $M5\perp M5\perp M5$-brane solution in M-theory, and the rotating black string solutions in five dimension, which relates the $M2\perp M2 \perp M2$-brane solution in M-theory. M-Kerr solution, which is $M5\perp M5\perp M5$-brane, becomes $NS5\perp D4\perp D4$-brane with kk-wave solutions in type IIA super string theory. M-BS solutions, which is $M2\perp M2\perp M2$-brane, are the same configuration of non-BPS black ring solution given by \cite{NONBPSBR}, and M-BS solution include the limit for $R\rightarrow \infty$ for the non-BPS two rotating charged black ring solution. 

We can apply these solutions to the AdS/CFT correspondence, $\alpha'\rightarrow 0$ with fixed $r/\alpha'$ for CFT, where $r$ is the distance from the D-brane. We will show the micro state of these solutions in the context of AdS/CFT correspondence, and we will also show the regular solutions with the specific physical parameters in subsequent paper, witch we are writing now. In the limit for the CFT, we compare the micro state of Kerr black hole by $D0\perp D6$-brane solutions given by Horowitz et. al., \cite{KerrMicro}. 

In this paper we do not apply the specific five dimensional vacuum black hole solutions. It is easy to apply for various solutions \cite{BR,2BR,BS,BdiRing}, however the difficulty for the calculation is just the Hodge dual \eqref{eqn:Hodge1} and \eqref{eqn:dual3}. The integrable condition must be necessary for the explicit representation of the ${\tilde D}_{ab}$, and in the below case the Hodge dual field becomes easier. When the metric of base space can be written in $\gamma_{ab}dx^adx^b=\Theta(x,y) (\Omega_\phi d\phi-\Omega_\psi d\psi)^2 $, the dual gauge field ${\tilde D}_{\phi\psi}$ must be vanishing, thus we only consider the ${\tilde D}_{ta}$ after all. 

By the way of this paper we only consider from the vacuum solutions, but adding the extra dimension we can extend to the Einstein manifold with the constant gauge field, which satisfy the Einstein and Maxwell equation in lower and higher dimension. In this formalism, including the one rotating black ring case \cite{NONBPSBR}, we can apply the more interesting case, especially cosmology and black hole dynamics.

\section*{Acknowledgements}
We wish to thank Nobuyoshi Ohta for useful discussions and Kei-ichi Maeda for the great suggestion for this work. We also wish to show our acknowledgement to the financial support by The 21st Century COE Program (Holistic Research and Education Center for Physics Self-organization Systems) at Waseda University. 





\begin{thebibliography}{99}

\bibitem{BR}R. Emparan,
[arXiv:hep-th/0402149].

\bibitem{2BR}
A. A. Pomeransky and R. A. Sen'kov, [arXiv:hep-th/0612005].

\bibitem{CBR}H. Elvang, 
Phys. Rev. D68 (2003) 124016, [arXiv:hep-th/0305247]. 

\bibitem{ST}H. Elvang and R. Emparan,
[arXiv:hep-th/0310008].

\bibitem{BS}H. Elvang and P. Figueras, 
[arXiv:hep-th/0701035]. 

\bibitem{BdiRing}H. Iguchi and T. Mishima, 
Phys. Rev. D 75 (2007) 0640018, [arXiv:hep-th/0701043]. 

\bibitem{BC}H. Elvang adn M. M. Rodriguez, 
[arXiv:0712.2425].

\bibitem{rod}R. Emparan and H. S. Reall, 
Phys.Rev. D65 (2002) 084025, [arXiv:hep-th/0110258].

\bibitem{SolGen}H. Iguchi and T. Mishima, 
Phys. Rev. D 74 (2006) 024029, [arXiv:hep-th/0605090].

\bibitem{BoostBrane}J. G. Russo and A. A. Tseytlin,
 Nucl. Phys.  B 490, 121 (1997),  [arXiv:hep-th/9611047].

\bibitem{BoundMBrane}N. Ohta and J. G. Zhou,
 Int. J. Mod. Phys.  A 13, 2013 (1998), [arXiv:hep-th/9706153].

\bibitem{Tdual}E. Bergshoeff, C. M. Hull and T. Ortin,
 Nucl. Phys. B 451, 547 (1995), [arXiv:hep-th/9504081].

\bibitem{Udual}M. Cvetic and C. M. Hull, 
Nucl. Phys. B 480, 296 (1996), [arXiv:hep-th/9606193]

\bibitem{NONBPSBR}H. Elvang, R Emparan and P. Figueras, 
JHEP 0502 (2005) 031, [arXiv:hep-th/0412130].

\bibitem{AdS/CFT}O. Aharony, S. S. Gubser, 
J. M. Maldacena, H. Ooguri and Y. Oz, 
Phys. Rept. 323, 183 (2000) [arXiv:hep-th/9905111]. 

\bibitem{GLinst}R. Gregory adn R. Laflamme, 
Phys. Rev. D 37 (1988) 305.

\bibitem{GLreview}T. Harmark, V. Niarchos and N. A. Obers,
Class. Quant. Grav. 24, R1 (2007), [arXiv:hep-th/0701022]. 

\bibitem{SUSYBR}H. Elvang, R. Emparan, D. Mateos and H. S. Reall, 
Phys. Rev. D71 (2005) 024033, [arXiv:hep-th/0408120]. 

\bibitem{Rot3Charge}M. Cvetic and D. Youm,
Nucl. Phys. B476 (1996) 118, [arXiv:hep-th/9612229].

\bibitem{SenBH}A. Sen, 
Phys.Rev.Lett. 69 (1992) 1006, [arXiv:hep-th/9204046].

\bibitem{SenBS}A. Sen, 
Nucl.Phys. B440 (1995) 421, [arXiv:hep-th/9411187].

\bibitem{GibbonsMaeda}, G. W. Gibbons and K. Maeda, 
Nucl. Phys. B 298 (1998) 741. 

\bibitem{smoothD1D5P}V. Jejjala, O. Madden, S. F. Ross and G. Titchener, 
Phys.Rev. D71 (2005) 124030, [arXiv:hep-th/0504181]. 

\bibitem{RotatingM}M. Cvetic and D Youm, 
Nucl.Phys. B499 (1997) 253, [arXiv:hep-th/9612229].

\bibitem{SSfromIM}K. Maeda and M Tanabe, 
Nucl. Phys. B738 (2006) 184, [arXiv:hep-th/0510082].

\bibitem{SUSYComp}K. Maeda, N. Ohta and M. Tanabe, 
Phys.Rev. D74 (2006) 104002, [arXiv:hep-th/0607084].

\bibitem{KerrMicro}G. T. horowitz and M. M. Roberts, 
[arXiv:0708.1346]. 


\end{thebibliography}
\end{document}